\documentclass[prd,twocolumn,floatfix,amsmath,nofootinbib,amssymb,floatfix]{revtex4}
\usepackage{graphicx,color,dcolumn,booktabs,bm,multirow}
\usepackage{longtable,lscape}
\usepackage{txfonts}
\usepackage{overpic}
\usepackage{amssymb}
\usepackage{array}
\usepackage{indentfirst}
\usepackage{feynmf}   
\usepackage{slashed}  
\usepackage{cases}
\usepackage{color}
\usepackage{multirow}
\usepackage{epstopdf}
\usepackage{tabularx}
\usepackage{graphicx,color,dcolumn,booktabs,bm}
\usepackage[compat=1.1.0]{tikz-feynman}
\usepackage[colorlinks,
citecolor=blue,
anchorcolor=red,
menucolor=red,
linkcolor=red,
filecolor=red,
runcolor=red,
urlcolor=blue,
frenchlinks=red]{hyperref}

\begin{document}

\title{$X(4630)$ and $Y(4626)$ production in the $B^+$ and $B_s^0$ decays}
\author{Zhuo Yu$^{1}$}\email{zhuo@seu.edu.cn}
\author{Qi Wu$^{2}$}\email{wuqi@htu.edu.cn}
\author{Dian-Yong Chen$^{1,3}$\footnote{Corresponding author}}\email{chendy@seu.edu.cn}
\affiliation{$^1$ School of Physics, Southeast University, Nanjing 210094, China\\
$^2$Institute of Particle and Nuclear Physics, Henan Normal University, Xinxiang 453007, China\\
$^3$ Lanzhou Center for Theoretical Physics, Lanzhou University, Lanzhou 730000, China}
\date{\today}

\begin{abstract}
In the present work, we investigate the production of $X(4630)$ and $Y(4626)$ in $B^+$ and $B_s^0$ decays, where $X(4630)$ and $Y(4626)$ are considered as the $C-$ parity pigeon pair in the $D_{s}^{\ast+} D_{s1}(2536)^-$ molecular frame. The branching fractions of $B^+ \to K^+ X(4630)/Y(4626)$ and $B_s^0 \to \eta X(4630)/Y(4626)$ have been evaluated using an effective Lagrangian approach, which are of the order of $10^{-5}$ and the ratios of these branching fractions are almost independent on the model parameter. Based on the present estimations, we propose to search $Y(4626)$ in the process $B^+ \to K^+ J/\psi \eta^{(\prime)}$, which should be accessible by the LHCb and Belle II Collaborations.                                             
\end{abstract}
\maketitle

\section{introduction}
\label{sec:introduction}
The observations of $X(3872)$ by the Belle Collaboration in 2003 served as a strong indication of the existence of exotic states~\cite{Belle:2003nnu}. Since then, an increasing number of exotic candidates have been reported by the LHCb, Belle, BaBar, and BES III Collaborations (see Refs.~\cite{Chen:2016qju,Hosaka:2016pey,Lebed:2016hpi,Esposito:2016noz,Guo:2017jvc,Ali:2017jda,Olsen:2017bmm,Karliner:2017qhf,Yuan:2018inv,Dong:2017gaw,Liu:2019zoy} for a recent review). Among these exotic candidates, charmonium-like states, typically observed in hidden charm decay processes, constitute the largest category. 

In 2016, the LHCb Collaboration performed the first full amplitude analysis of $B^+ \to K^+ J/\psi \phi $ decays was performed with a data sample of $3\ \mathrm{fb}^{-1}$ of $pp$ collision data collected in Run I~\cite{LHCb:2016axx, LHCb:2016nsl}, four charmonium-like states, $X(4140)$, $X(4274)$, $X(4500)$ and $X(4700)$ were reported in the $J/\psi \phi$ invariant mass distribution with the significance exceeding 5 standard deviations. Later on, they reanalysised the same process using the proton-proton collision data corresponding to a total integrated luminosity of $9\ \mathrm{fb}^{-1}$~\cite{LHCb:2021uow}, four previously reported charmonium-like states in the $J/\psi \phi$ invariant mass distributions have been confirmed, and to well describe the distributions of $\phi K^+$, $J/\psi \phi$, and $J/\psi K^+$ invariant mass for the $B^+ \to K^+ J/\psi \phi $, two new structures, $Z_{cs}(4000)$ and $Z_{cs}(4220)$, in the $J/\psi K$ invariant mass distributions, and three additional states, $X(4150)$, $X(4630)$, and $X(4685)$ in the $J/\psi \phi$ invariant mass distributions were introduced. The resonance parameters of $X(4630)$ were reported to be~\cite{LHCb:2021uow},
\begin{eqnarray}
\begin{split}
	m_X &= \left(4626\pm16^{+18}_{-110}\right) \, {\rm MeV}, \\
    \Gamma_X &= \left(174\pm27^{+134}_{-73}\right) \, {\rm MeV},
\end{split}
\end{eqnarray}
respectively, and its spin-parity were measured to be $1^-$ with a significance 5.5 $\sigma$. In addition, the $C$-parity of $X(4630)$ must be positive since it is observed in the $J/\psi \phi$ final states. Thus, it is interesting to notice that $X(4630)$ should be an exotic state since quantum numbers $J^{PC}=1^{-+}$ are not allowed by conventional meson states.

Besides the resonance parameters, the LHCb Collaboration also reported the fit fraction of $X(4630)$ in the process $B^+ \rightarrow J/\psi \phi K^+$, which is~\cite{LHCb:2021uow},
\begin{eqnarray}
	\frac{\mathcal{B}[B^+ \rightarrow K^+ X(4630)\rightarrow  K^+ J/\psi \phi]}{\mathcal{B}[B^+ \rightarrow K^+ J/\psi \phi ]}=\left(2.6\pm0.5_{-1.5}^{+2.9}\right)\%.
\end{eqnarray}
The branching ratio of $B^+ \rightarrow K^+ J/\psi \phi $ is $(5.0\pm 0.4)\times10^{-5}$ \cite{ParticleDataGroup:2020ssz}, thus, we conclude that the branching fraction of the cascade process should be,
\begin{eqnarray}
\mathcal{B}[B^+ \rightarrow K^+X(4630)\rightarrow K^+ J/\psi \phi ]=\left(1.3^{+1.5}_{-0.8}\right)\times10^{-6}.
\end{eqnarray} 

In addition to $X(4630)$, there is another interesting charmonium-like state in the vicinity of 4.6 GeV, which is $Y(4626)$, observed in the cross sections for $e^+e^- \to D_s^{\ast +} D_{s1}^-+c.c.$ by the Belle Collaborations in 2019~\cite{Belle:2019qoi}. The measured mass and width are,
\begin{eqnarray}
\begin{split}
	m_Y &= \left(4625.9^{+6.2}_{-6.0}{\rm (stat.)}\pm0.4{\rm (syst)} \right)\, {\rm MeV}, \\
	\Gamma_Y &=  \left(49.8^{+13.9}_{-11.5}{\rm (stat.)}\pm4.0{\rm (syst)} \right)\, {\rm MeV},
\end{split}
\end{eqnarray} 
respectively. The $J^{PC}$ quantum numbers of $Y(4626)$ are $1^{--}$ since it is observed in the $e^+e^-$ annihilation process. The mass of this states is almost the same as that of $X(4630)$ but with different $C$-parity. Thus, it is interesting to investigate their properties simultaneously. In the tetraquark scenario, the authors in Refs.~\cite{Deng:2019dbg} utilized a multiquark color flux tube model with a multibody confinement potential and a one-gluon-exchange interaction to make an exhaustive investigation on the diquark-antidiquark state $[cs][\bar{c}\bar{s}]$, and the estimations indicated that $Y(4626)$ was a $P$-wave $[cs][\bar{c}\bar{s}]$ tetraquark state. Using the QCD sum rule, the authors assigned $Y(4626)$ as a $P$-wave $[cs][\bar{c} \bar{s}]$ tetraquark system~\cite{Zhang:2020gtx}, and $X(4630)$ as the $[sc]_S[\bar{s}\bar{c}]_{\tilde{V}}+[sc]_{\tilde{V}}[\bar{s}\bar{c}]_V$ tetraquark state~\cite{Wang:2023jaw}. The mass of $X(4630)$ could be reproduced in the diquark-antidiquark teraquark state by using the QCD two-point sum rule method~\cite{Agaev:2022iha}. While the estimations in the chiral constituent quark model with the help of the Gaussian expansion method found that $Y(4620)$ could be a resonance state~\cite{Tan:2019knr}.

Another interesting phenomenon for $X(4630)$ and $Y(4626)$ is that their masses are very close to the threshold of $D_s^{\ast+} {D}_{s1}(2536)^-$, which simulates theorists' great interest in the molecular interpretations of these two states. In fact, before the observation of LHCb Collaboration, $D_s^\ast \bar{D}_{s1}-D_{s1}\bar{D}_s^\ast$ molecular state with $J^{PC}=1^{-+}$ has been predicted~\cite{Wang:2020cme}, which is well consistent with $X(4630)$. In Ref.~\cite{Yang:2021sue}, the estimations of the one-boson exchange model indicated that $X(4630)$ could be assigned as a $D_s^{\ast +} D_{s1}(2536)^-$ molecular state with $J^{PC}=1^{-+}$ and the quasipotential Bethe-Salpeter equation calculations with one-boson exchange model suggested $Y(4626)$ as a $D_s^{\ast+} D_{s1}(2536)^-$ molecular state with $J^{PC}=1^{--}$~\cite{He:2019csk}. Using the QCD sum rule method, the authors of Ref.~\cite{Wang:2021ghk} tentatively assigned X(4630) as the $D_s^{\ast +} {D}_{s1}(2536)^-$ molecular molecular state or the $[cs]_P[\bar{c}\bar{s}]_A+[cs]_A[\bar{c}\bar{s}]_P$ tetraquark state with the $J^{PC}=1^{-+}$. Estimates using phenomenological models, such as the one-boson exchange model and the renormalization group saturation model, support $X(4630)$ and $Y(4626)$ to be $D_s^{\ast +} {D}_{s1}(2536)^-$ molecules with $J^{PC}=1^{-+}$ and $1^{--}$, respectively~\cite{Peng:2022nrj}.

Besides the resonance parameters, the production properties are also essential for revealing the inner structure of these charmonium-like states. As indicated in the above literature, $X(4630)$ and $Y(4626)$ could be $D_s^{\ast +} {D}_{s1}(2536)^-$ molecules with $J^{PC}=1^{-+}$ and $1^{--}$, respectively. In the present work, we investigate the production process $B^+ \to K^+ X(4630)$, which has been observed by LHCb Collaboration. With the same production mechanism, we can predict the production process $B^+ \to K^+ Y(4626)$. Moreover, the production processes $B_s^0 \to \eta X(4630)$ and $B_s^0 \to \eta Y(4626)$ can be also investigate in the similar production mechanism. All the production processes estimated in the present work should be accessible for further experimental measurements by LHCb and Belle II Collaborations.

This work is organized as follows. After the introduction, we present our estimations of branching fractions in Sec~\ref{Sec:Method}. The numerical results and relevant discussion are given in Sec~\ref{Sec:Num} and the last section is devoted to a short summary.

\section{Theoretical framework}
\label{Sec:Method}

In the present work, we consider both $X(4630)$ and $Y(4626)$ as $S$-wave molecular states composed of $D_s^\ast \bar{D}_{s1}$ with different $C$-party. Here we employ an effective Lagrangian approach to describe the coupling between the molecule and its components, and the coupling of $X(4630)/Y(4626)$ and $D_s^\ast \bar{D}_{s1}$ read,
\begin{eqnarray}
	{\cal L}_{X}&=&\frac {g_{X}}{\sqrt{2}}\epsilon_{\mu\nu\alpha\beta} \partial^{\mu} X^{\dagger\nu}\left( D_{s1}^{+\alpha} {D}_s^{*-\beta}-D_s^{*+\alpha} {D}_{s1}^{-\beta} \right ) ,\nonumber\\
	{\cal L}_{Y}&=&\frac {g_{Y}}{\sqrt{2}}\epsilon_{\mu\nu\alpha\beta} \partial^{\mu} X^{\dagger\nu}\left( D_{s1}^{+\alpha} {D}_s^{*-\beta}+D_s^{*+\alpha}{D}_{s1}^{-\beta} \right ),
	\label{Eq:LagMol}
\end{eqnarray}
with $g_X$ and $g_Y$ to be coupling constants, which  would be discussed later.

\begin{figure}[t]
 \begin{tabular}{cccc}
		\centering
		\includegraphics[width=4cm]{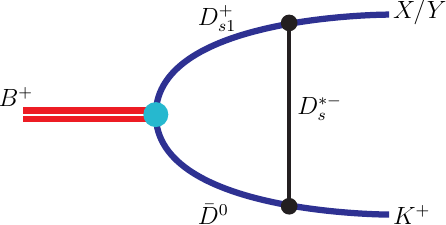}&
		\includegraphics[width=4cm]{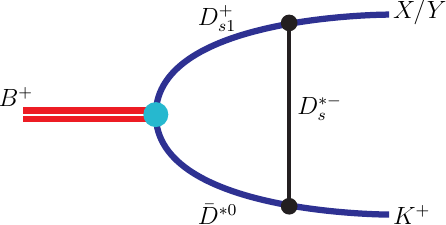}\\
		$(a)$ & $(b)$ \\ \\
		\includegraphics[width=4cm]{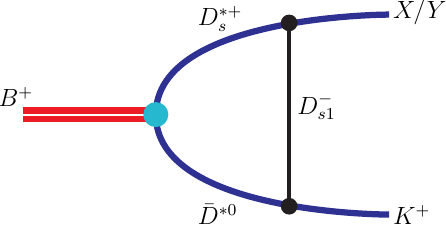}&
		\includegraphics[width=4cm]{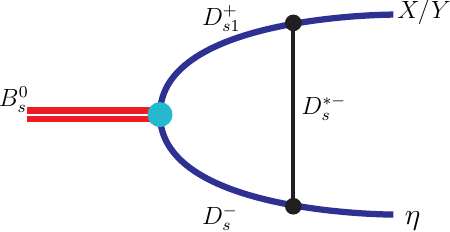}\\
		$(c)$ & $(d)$ \\ \\
        \includegraphics[width=4cm]{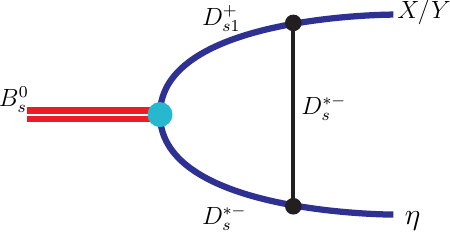}&
		\includegraphics[width=4cm]{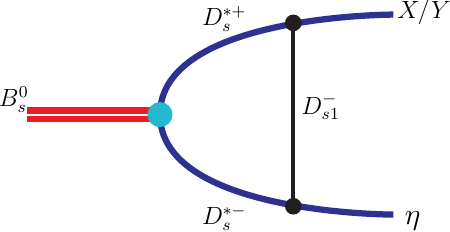}&\\
		\\
		$(e)$  & $(f)$ \\
 \end{tabular}
  \caption{Diagrams contributing to $B^+ \to K^+ X(4630)/Y(4626)$ (diagrams (a)-(c)) and $B^0_s\rightarrow \eta X(4630)/Y(4626)$ (digrams (e)-(f)).} \label{Fig:Mech}
\end{figure}

\subsection{Production Mechanism}
Here we take decay process $B^+\to K^+ X(4630)$ as an example to discuss the production mechanism. At the quark level, this process could be divided into two parts. First, the anti-bottom quark in the bottom meson decays into an anti-charm quark by emitting a $W^+$ boson, then the anti-charm quark and the spectated $u$ quark can hadronized as an anti-charmed meson. As for the $W^+$ boson, it transits into the $c\bar{s}$ quark pair, which could hadronize as a charmed-strange meson subsequently. Then, the anti-charmed and charmed strange mesons couple to $X(4630)$ and $K^+$ by exchanging a proper meson state. Here, we use $\mathcal{H}_W$ and $\mathcal{H}_T$ to refer the first weak interaction process and the consequent transition process, respectively, then, the decay process could be expressed as,
\begin{eqnarray}
&& \langle K^+ X(4630) \left|\mathcal{H}_T \mathcal{H}_W \right|B^+\rangle \nonumber\\ 
&&\qquad=\langle K^+ X(4630) \left|\mathcal{H}_T \right|(\bar{c}u)(c\bar{s})\rangle \langle (\bar{c}u)(c\bar{s}) \left|  \mathcal{H}_W \right|B^+\rangle \ \ \ 
\end{eqnarray}
In principle, all the possible anti-charmed and charmed-strange mesons, which could connect the initial $B^+$ and final $K^+ X(4630)$ should be considered. However, in the molecular frame, $X(4630)$ strongly couples to its components $D_s^{\ast +} {D}_{s1}(2536)^-$, and consequently, the diagrams in Fig.~\ref{Fig:Mech}-(a)-(c) should contribute dominantly to the process $B^+\to K^+ X(4630)$. Considering that both $X(4630)$ and $Y(4626)$ are $D_s^{\ast+} {D}_{s1}^{-}$ molecules, we can find that the production mechanism of $B^+ \to K^+ Y(4626)$ is the same as that of $B^+\to K^+ X(4630)$. Similarly, we can construct the productions of $X(4630)/Y(4626)$ in the $B_s^0$ decays, where the spectated $u$ quark is replaced by a $s$ quark, and consequently the final $K^+$ should become a $\eta$ meson. The corresponding diagrams contributing to $B_s^0 \to \eta X(4630)/Y(4626)$ are listed in Fig.~\ref{Fig:Mech}-(d)-(f). Such a mechanism has been widely employed to investigate the productions of molecular states in bottom and bottom strange hadrons decays~\cite{Yu:2023avh, Wu:2024lud, Wu:2023fyh, Wu:2021cyc, Wu:2021caw, Wu:2019rog}.

\subsection{Effective Lagrangian}
In the present work, all the diagrams listed in  Fig.~\ref{Fig:Mech}  are evaluated at the hadronic level and the interactions between the relevant hadrons are described by effective Lagrangians. As for $B$ and $B_s$ weak decay,  we utilize the parametrized hadronic matrix elements obtained by the effective Hamiltonian at the quark level, which are~\cite{Cheng:2003sm,Soni:2021fky}, 
\begin{eqnarray}
\label{eq:weak transition}
&&\left\langle0|J_\mu|P(p_1)\right\rangle = -if_p p_{1\mu}\;,\nonumber\\
&&\left\langle0|J_\mu|V(p_1,\epsilon)\right\rangle =f_V \epsilon_\mu m_V \;,\nonumber\\
&&\left\langle P(p_2)|J_\mu|B(p)\right\rangle \nonumber\\&&\quad=\left[P_\mu-\frac{m^2_{B}-m^2_P}{q^2}q_\mu\right]F_1\left(q^2\right) 
+\frac{m^2_{B}-m^2_P}{q^2}q_\mu F_0\left(q^2\right) \;,\nonumber\\
&&\left\langle V(p_2,\epsilon)|J_\mu|B(p)\right\rangle \nonumber\\
&&\quad=\frac{i\epsilon^\nu}{m_{B}+m_V}\Bigg\{i\varepsilon_{\mu\nu\alpha\beta}P^\alpha q^\beta A_V\left(q^2\right)\nonumber\\
&&\quad +(m_{B}+m_V)^2 g_{\mu\nu}A_1\left(q^2\right) -P_\mu P_\nu A_2\left(q^2\right)\nonumber\\
&&\quad-2m_V(m_{B}+m_V)\frac{P_\nu q_\mu}{q^2}\left[A_3\left(q^2\right)-A_0\left(q^2\right)\right] \Bigg\}\;,\nonumber\\
&&\left\langle P(p_2)|J_\mu|B_{s}(p)\right\rangle = F_+\left(q^2\right)P^\mu+F_-\left(q^2\right)q^\mu\;,\nonumber\\
&&\left\langle V(p_2,\epsilon)|J_\mu|B_{s}(p)\right\rangle\nonumber\\
&&\quad =\frac{\epsilon_\nu}{m+m_2}\Bigg[-g^{\mu\nu}P\cdot qA_0\left(q^2\right)+P^\mu P^\nu A_+\left(q^2\right)\nonumber\\
&&\qquad+q^\mu P^\nu\left(q^2\right)A_-\left(q^2\right)+i\varepsilon^{\mu\nu\alpha\beta}P_\alpha q_\beta V\left(q^2\right) \Bigg]\;,
	\end{eqnarray}
with $J_{\mu}=\bar{q}_1 \gamma_{\mu}(1-\gamma5)q_2$, $P_{\mu}=p_\mu+p_{2\mu}$ and $q_{\mu}=p_{\mu}-p_{2\mu}$.  $A_{0,1,2,V,+,-}\left(q^2\right)$ and $F_{0,1,+,-}\left(q^2\right)$ are the weak transition form factors, while $A_3\left(q^2\right)$ is the linear combination of form factors $A_1\left(q^2\right)$ and $A_2\left(q^2\right)$, which is~\cite{Cheng:2003sm},
\begin{eqnarray}
A_3\left(q^2\right)=\frac{m_B+m_{V}}{2m_{V}}A_1\left(q^2\right)-\frac{m_B-m_{V}}{2m_{V}}A_2\left(q^2\right). 
\end{eqnarray}

With the hadronic matrix elements above, we can get the amplitude of the first weak transition vertex by the products of two hadronic matrix elements. Here we take  the process of $B^+ \rightarrow D_{s1}^+\bar{D}^*$ as an example, the amplitude reads,
\begin{eqnarray}
&&\mathcal{A}(B^+ \to D_{s1}^+ \bar{D}^*) =	\mathcal{A}_{\mu\nu }^{B^+ \to D_{s1}^+ \bar{D}^*} \epsilon_{D_{s1}}^\mu \epsilon_{\bar{D}^\ast}^\nu \nonumber\\
&&\qquad = \frac{G_F}{\sqrt{2}}V_{cb}V_{cs}a_1	\left \langle D_{s1}^+ \left | J_\mu \right |0  \right \rangle 	\left \langle \bar{D}^* \left | J_\mu \right |B^+  \right \rangle
\end{eqnarray}
where $G_F$ is the Fermi constant,$V_{cb}$ and $V_{cs}$ are the CKM matrix elements. $a_1=c_1^{eff}+c_2^{eff}/N_c$ with $c_{1,2}^{eff}$ are effective Wilson coefficients obtained by the factorization approach\cite{Bauer:1986bm}. In the present work, we adopt  $G_F=1.166 \times 10^{-5} {\rm GeV}^{-2}$, $V_{cb}=0.041$,$V_{cs}=0.987$ and $a_1=1.05$ as in Refs.~\cite{ParticleDataGroup:2020ssz,Ali:1998eb,Ivanov:2006ni}.

Besides the effective Lagrangian related to the molecular states and the weak interactions discussed above, the effective Lagrangian for the light pseudoscalar states and heavy-light mesons   are also needed, which could be constructed based on heavy quark limit and chiral symmetry~\cite{Casalbuoni:1996pg,Casalbuoni:1992gi,Casalbuoni:1992dx}, and the relevant effective Lagrangians read,
\begin{eqnarray}
\label{eq:D(1)DP}
\mathcal{L}_{\mathcal{D}\mathcal{D}^*\mathcal{P}}& =& g_{\mathcal{D}\mathcal{D}^*\mathcal{P}}\mathcal{D}_b\left(\partial_{\mu} \mathcal{P}\right)_{ba}\mathcal{D}_a^{*\mu\dagger}\nonumber\\
	\mathcal{L}_{\mathcal{D}^*\mathcal{D}^*\mathcal{P}} &=& g_{\mathcal{D}^*\mathcal{D}^*\mathcal{P}}\varepsilon_{\mu\nu\alpha\beta}\left(\mathcal{D}_b^{*\mu}\overleftrightarrow{\partial^{\alpha}}\mathcal{D}_a^{*\nu\dagger}\right) \partial^{\beta}\mathcal{P}_{ba}\nonumber\\
	\mathcal{L}_{\mathcal{D}^*\mathcal{D}_1\mathcal{P}}&=& g_{\mathcal{D}^*\mathcal{D}_1\mathcal{P}}\Big[3\mathcal{D}_{1b}^{\mu}\left(\partial_{\mu}\partial_{\nu}\mathcal{P}\right)_{ba}\mathcal{D}_a^{*\nu\dagger}-\nonumber\\ &&\mathcal{D}_{1b}^{\mu}\left(\partial^{\nu}\partial_{\nu}\mathcal{P}\right)_{ba}\mathcal{D}_{a\mu}^{a\dagger}\Big],
\end{eqnarray}
with $\mathcal{D}=(D^0, D^+, D_s^+)$ and $\mathcal{D}_1=(D_1^0, D_1^+, D_{s1}^+)$, $\mathcal{P}$ is the matrix form of the light pseudoscalar meson with concrete form as,
\begin{eqnarray}\label{eq:pscalar matrix}
	\mathcal{P} &=&
	\left(\begin{array}{ccc}
		\frac{\pi^{0}}{\sqrt 2}+\alpha\eta+\beta\eta^\prime &\pi^{+} &K^{+}\\
		\pi^{-} &-\frac{\pi^{0}}{\sqrt2}+\alpha\eta+\beta\eta^\prime&K^{0}\\
		K^{-} &\bar K^{0} &\gamma\eta+\delta\eta^\prime
	\end{array}\right)
\end{eqnarray}
where the parameters $\alpha$, $\beta$, $\gamma$ and $\delta$ related to the mixing angle $\theta$ are defined as
\begin{eqnarray}
	\alpha&=&\frac{\cos\theta-\sqrt{2}\sin\theta}{\sqrt{6}},\ \quad \beta=\frac{\sin\theta+\sqrt{2}\cos\theta}{\sqrt{6}},\nonumber\\
	\gamma&=&\frac{-2\cos\theta-\sqrt{2}\sin\theta}{\sqrt{6}},\ \delta=\frac{-2\sin\theta+\sqrt{2}\cos\theta}{\sqrt{6}},
\end{eqnarray}
with the mixing angle $\theta=-19.1^\circ$ \cite{MARK-III:1988crp,DM2:1988bfq}. It is worth noting that only $S$-wave couplings are considered in Eq.~\eqref{eq:D(1)DP}.

\subsection{Decay Amplitude}
With the effective Lagrangians discussed in above subsection, we can now obtain the amplitudes of $B^+ \rightarrow K^++X(4630)$ corresponding to Fig.~\ref{Fig:Mech}-(a)-(c),which are,
\begin{eqnarray}\label{eq:amp}
			\mathcal{M}_{a}&=&  \int\frac{d^4q}{(2\pi)^4}\Big[\mathcal{A}^{B\rightarrow D_{s1} \bar{D}^{0}}_{\mu}(p_1,p_2)\Big]\Big[-\frac{g_{X}}{\sqrt{2}}\varepsilon_{\mu_1\theta\alpha\beta}p_3^{\mu_1}\epsilon^{\theta}(p_3)\Big]\nonumber\\
			&&\times \Big[-g_{DD_s^*K}p_{4\sigma}\Big] \Big[\frac{-g^{\mu\alpha}+p_1^{\mu} p_1^{\alpha}/m_1^2}{p_1^2-m_1^2}\Big]\Big[\frac{1}{p_2^2-m_2^2}\Big] \nonumber\\
			&&\times\Big[\frac{-g^{\beta\sigma}+q^{\beta} q^{\sigma}/m_q^2}{q^2-m_q^2}\Big]\mathcal{F}\left(q^2,m_q^2 \right)\nonumber
\end{eqnarray}
\begin{eqnarray}			
	\mathcal{M}_{b}&=& \int\frac{d^4q}{(2\pi)^4}\Big[\mathcal{A}^{B\rightarrow D_{s1} \bar{D}^{*0}}_{\mu\nu}(p_1,p_2)\Big]\Big[-\frac{g_{X}}{\sqrt{2}}\varepsilon_{\mu_1\theta\alpha\beta}p_3^{\mu_1}\epsilon^{\theta}(p_3)\Big]\nonumber\\
			&&\times\Big[-g_{D^*D_s^*K}\varepsilon_{\rho\sigma\alpha_2\beta_2}(q+p2)^{\alpha_2}p_4^{\beta_2}\Big] \Big[\frac{-g^{\mu\alpha}+p_1^{\mu} p_1^{\alpha}/m_1^2}{p_1^2-m_1^2}\Big] \nonumber\\
			&&\times\Big[\frac{-g^{\nu\rho}+p_2^{\nu} p_2^{\rho}/m_2^2}{p_2^2-m_2^2}\Big]\Big[\frac{-g^{\beta\sigma}+q^{\beta} q^{\sigma}/m_q^2}{q^2-m_q^2}\Big]\mathcal{F}\left(q^2,m_q^2 \right)\nonumber\\
	\mathcal{M}_{c}&=& \int\frac{d^4q}{(2\pi)^4}\Big[\mathcal{A}^{B\rightarrow D_{s}^* \bar{D}^{*0}}_{\mu\nu}(p_1,p_2)\Big]\Big[-\frac{g_{X}}{\sqrt{2}}\varepsilon_{\mu_1\theta\alpha\beta}p_3^{\mu_1}\epsilon^{\theta}(p_3)\Big]\nonumber\\
			&&\times \Big[-g_{D^*D_{s1}K}\left( g_{\rho\sigma}p_4^{2}-3p_{4\rho}p_{4\sigma}\right)\Big]\Big[\frac{-g^{\mu\alpha}+p_1^{\mu} p_1^{\alpha}/m_1^2}{p_1^2-m_1^2}\Big]\nonumber\\
			&&\times\Big[\frac{-g^{\nu\rho}+p_2^{\nu} p_2^{\rho}/m_2^2}{p_2^2-m_2^2}\Big]\Big[\frac{-g^{\beta\sigma}+q^{\beta} q^{\sigma}/m_q^2}{q^2-m_q^2}\Big]\mathcal{F}\left(q^2,m_q^2 \right).\nonumber\\
\end{eqnarray}
Then the amplitude for $B^+\to K^+ X(4630)$ reads,
\begin{eqnarray}
	\mathcal{M}_{B\to K^+X} =\mathcal{M}_a+\mathcal{M}_b+\mathcal{M}_c. \label{Eq:AmpK}
\end{eqnarray}

In the same way, we can obtain the amplitudes for $B_s^0 \to \eta X(4630)$ corresponding diagrams Fig.~\ref{Fig:Mech}-(d)-(f), which are, 
\begin{eqnarray}			
		\mathcal{M}_{d}&=&  \int\frac{d^4q}{(2\pi)^4}\Big[\mathcal{A}^{B_s\rightarrow D_{s1} \bar{D_s}}_{\mu}(p_1,p_2)\Big]\Big[-\frac{g_{X}}{\sqrt{2}}\varepsilon_{\mu_1\theta\alpha\beta}p_3^{\mu_1}\epsilon^{\theta}(p_3)\Big]\nonumber\\
			&&\times
		\Big[-g_{D_sD_s^*\eta}p_{4\sigma}\Big]\Big[\frac{-g^{\mu\alpha}+p_1^{\mu} p_1^{\alpha}/m_1^2}{p_1^2-m_1^2}\Big]\Big[\frac{1}{p_2^2-m_2^2}\Big]\nonumber\\
			&&\times\Big[\frac{-g^{\beta\sigma}+q^{\beta} q^{\sigma}/m_q^2}{q^2-m_q^2}\Big]\mathcal{F}\left(q^2,m_q^2 \right)\nonumber\\
	\mathcal{M}_{e}&=&  \int\frac{d^4q}{(2\pi)^4}\Big[\mathcal{A}^{B_s\rightarrow D_{s1} \bar{D_s}^*}_{\mu}(p_1,p_2)\Big]\Big[-\frac{g_{X}}{\sqrt{2}}\varepsilon_{\mu_1\theta\alpha\beta}p_3^{\mu_1}\epsilon^{\theta}(p_3)\Big] \nonumber\\
			&&\times\Big[- g_{D_s^*D_s^*\eta}\varepsilon_{\rho\sigma\alpha_2\beta_2}(q+p_2)^{\alpha_2}p_4^{\beta_2}\Big] \Big[\frac{-g^{\mu\alpha}+p_1^{\mu} p_1^{\alpha}/m_1^2}{p_1^2-m_1^2}\Big]\nonumber\\
			&&\times\Big[\frac{-g^{\nu\rho}+p_2^{\nu} p_2^{\rho}/m_2^2}{p_2^2-m_2^2}\Big]\Big[\frac{-g^{\beta\sigma}+q^{\beta} q^{\sigma}/m_q^2}{q^2-m_q^2}\Big]\mathcal{F}\left(q^2,m_q^2 \right)\nonumber\\
	\mathcal{M}_{f}&=&  \int\frac{d^4q}{(2\pi)^4}\Big[\mathcal{A}^{B_s\rightarrow D_{s}^* \bar{D_s}^*}_{\mu}(p_1,p_2)\Big]\Big[-\frac{g_{X}}{\sqrt{2}}\varepsilon_{\mu_1\theta\alpha\beta}p_3^{\mu_1}\epsilon^{\theta}(p_3)\Big]\nonumber\\
			&&\times\Big[i~g_{D_s^*D_{s1}\eta}\left(g_{\rho\sigma} p_4^{\nu_2}p_{4{\nu_2}}-3p_{4\rho}p_{4\sigma} \right)\Big]\nonumber\\
			&&\times \Big[\frac{-g^{\mu\alpha}+p_1^{\mu} p_1^{\alpha}/m_1^2}{p_1^2-m_1^2}\Big]\Big[\frac{-g^{\nu\rho}+p_2^{\nu} p_2^{\rho}/m_2^2}{p_2^2-m_2^2}\Big]\nonumber\\
			&&\times\Big[\frac{-g^{\beta\sigma}+q^{\beta} q^{\sigma}/m_q^2}{q^2-m_q^2}\Big]\mathcal{F}\left(q^2,m_q^2 \right),	
\end{eqnarray}
then the amplitude for $B_s^0 \to \eta X(4630)$ reads,
\begin{eqnarray}
	\mathcal{M}_{B_s^0 \to \eta X} =\mathcal{M}_d +\mathcal{M}_e +\mathcal{M}_f.\label{Eq:AmpEta}
\end{eqnarray}

In the the above amplitudes, we introduce a form factor $\mathcal{F}(q^2,m^2)$ in monopole form to compensate for the off-shell effect and avoid the ultraviolet divergence in the loop integral. Its concrete form is,
\begin{equation}
	\mathcal{F}\left(q^2,m^2\right) = \frac{m^2-\Lambda^2}{q^2-\Lambda^2},
\end{equation}
where the parameter $\Lambda$ could be reparameterized as ${ \Lambda} = m + \alpha {\rm \Lambda_{QCD}}$ with $ \Lambda_{\rm QCD}$ = 220 MeV and $\alpha$ to be the model parameter~\cite{Cheng:2004ru}, which should be of order unity. 

With above amplitudes in Eqs.~(\ref{Eq:AmpK}) and (\ref{Eq:AmpEta}), the partial width can be estimated by,
\begin{eqnarray}
	\Gamma_{B^+\to K^+ X} &=& \frac{1}{8\pi} \frac{|\vec{p}\,|}{m_{B^+}^2}\overline{\left|\mathcal{M}_{B^+\to K^+ X}\right|^2},\nonumber\\
	\Gamma_{B_s^0\to \eta X} &=& \frac{1}{8\pi} \frac{|\vec{p}\,|}{m_{B_s^0}^2}\overline{\left|\mathcal{M}_{B_s^0\to \eta X}\right|^2}.
\end{eqnarray}
In the same manner, we can obtain the amplitudes for $B^+ \to K^+ Y(4626)$ and $B_s^0 \to \eta Y(4626)$, and then estimated the corresponding partial decay widths.


\section{Numerical Results and Discussions}
\label{Sec:Num}
\subsection{Coupling Constants}
Before we estimate the branching ratio of the relevant decay processes, some coupling constants should be further clarified. The coupling constants in Eq.~(\ref{Eq:LagMol}) could be estimated by Weinberg's compositeness theorem\cite{Weinberg:1965zz,Baru:2003qq,Chen:2017abq}, which is 
\begin{eqnarray}
 g_{X/Y}^2=16\pi \frac{(m_{D_s^*}+m_{D_{s1}})^{5/2}}{m_{X/Y}^2}\sqrt{\frac{2E_b}{m_{D_s^*}m_{D_{s1}}}}
\end{eqnarray}
by taking the center value of $X(4630)/Y(4626)$, we have $g_X \simeq g_Y \simeq 3.12~{\rm GeV^2}$.

Considering the chiral symmetry and heavy quark limit, the coupling constants of charmed meson with pseudoscalar meson can be related to some gauge coupling by~\cite{Isola:2003fh,Falk:1992cx, Casalbuoni:1996pg},
\begin{eqnarray}
	g_{DD^*P}&=&-g_{\bar{D}\bar{D^*}P}=-\frac{2g}{f_{\pi}}\sqrt{m_{D}m_{D^*}}\nonumber\\
	g_{D^*D^*P}&=&\frac{2g}{f_{\pi}}\frac{1}{\sqrt{m_{D^*}m_{D^*}}}\nonumber\\
	g_{D^*D_1P}&=&g_{\bar{D}^*\bar{D}_1P}=-\frac{\sqrt{6}}{3}\frac{h_1+h_2}{\Lambda_{\chi}f_{\pi}}\sqrt{m_{D^*}m_{D_1}}
\end{eqnarray}
with $g=0.59$, $f_{\pi}=0.132~{\rm GeV}$~\cite{Chen:2019asm,Liu:2011xc,Isola:2003fh,Falk:1992cx} and $(h_1+h_2)/\Lambda_{\chi}=0.55~{\rm GeV^{-1}}$~\cite{Casalbuoni:1996pg}.

\begin{table}[tb]
	\caption{The values of the parameters $F(0)$, $a$, and $b$ in the weak transition form factors for $B\rightarrow D^{(*)}$~\cite{Cheng:2003sm} and $B_s\rightarrow D^{(*)}_s$~\cite{Soni:2021fky}.}
	\label{Tab:F0ab}
	\renewcommand\arraystretch{1.5}
	\begin{tabular}{p{0.85cm}<\centering p{0.85cm}<\centering p{0.85cm}<\centering p{0.95cm}<\centering p{0.85cm}<\centering p{0.95cm}<\centering p{0.85cm}<\centering p{0.85cm}<\centering}
		\toprule[1pt]
		$F$ & $F(0)$ & $a$ & $b$ & $F$ & $F(0)$ & $a$ & $b$ \\
		\midrule[1pt]
		$F_0$ & 0.67 & 0.65 & 0.00 & $F_1$ & 0.67 & 1.25 & 0.39 \\
		$A_V$ & 0.75 & 1.29 & 0.45 & $A_0$ & 0.64 & 1.30 & 0.31 \\
		$A_1$ & 0.63 & 0.65 & 0.02 & $A_2$ & 0.61 & 1.14 & 0.52 \\ \midrule[1pt]
		$F_+$ & 0.770 & 0.837 & 0.077 & $F_-$ & -0.355 & 0.855 & 0.083 \\
		$A_+$ & 0.630 & 0.972 & -0.092 & $A_-$ & -0.756 &  1.001 &  0.116 \\
		$A_0$ & 1.564 & 0.442 & -0.178 & $V$ & 0.743 & 1.010 & 0.118 \\
		\bottomrule[1pt]
	\end{tabular}
\end{table}
\begin{table}[tb]
	\caption{Values of the parameters $\Lambda_1$ and $\Lambda_2$ for each weak transition form factors obtained by fitting Eq.~\eqref{eq:FQlambda} with Eq.~\eqref{eq:FQzeta}.}
	\label{Tab:L1L2}
	\renewcommand\arraystretch{1.5}
	\begin{tabular}{p{1.3cm}<\centering p{1.3cm}<\centering p{0.85cm}<\centering p{0.85cm}<\centering p{0.85cm}<\centering p{0.85cm}<\centering p{0.85cm}<\centering p{0.85cm}<\centering}
		\toprule[1pt]			
		Process& Parameter&   $A_V$ & $A_0$& $A_1$ & $A_2$ &   $F_0$ & $F_1$ \\
		\midrule[1pt]
		\multirow{2}{*}{$B\rightarrow D^{(*)}$} &$\Lambda_1$ & 6.32 & 5.32 & 7.83   & 7.35 &   7.75 & 6.53 \\
		&$\Lambda_2$ & 7.00& 9.41  & 10.99 & 7.35 &  11.00 &  6.84     \\
		\midrule[1pt]
		Process& Parameter&   $A_+$ & $A_-$ & $A_0$ & $V$&  $F_+$ &$F_1$\\ \midrule[1pt]
		\multirow{2}{*}{$B_s\rightarrow D_s^{(*)}$}    &$\Lambda_1$ & 5.48   & 5.77 & 9.75   & 5.74 &   6.30 & 6.25 \\
		&$\Lambda_2$ & 18.00  & 14.63  & 11.00 & 14.61 &  15.92 &  15.58    \\
		\bottomrule[1pt]
	\end{tabular}
\end{table}

As for the form factor in the weak transition amplitudes, they were parameterized in the form~\cite{Cheng:2003sm,Soni:2021fky}  
\begin{equation}\label{eq:FQzeta}
	F\left(q^2\right) = \frac{F(0)}{1-a\zeta+b\zeta^2}
\end{equation}
with $\zeta = Q^2/m_{B}^2$. The relevant parameters $F(0)$, $a$, and $b$ for each form factor are collected in Table~\ref{Tab:F0ab}. In order to simplify the estimations, we further parameterize these form factors in the form,
\begin{equation}\label{eq:FQlambda}
	F\left(q^2\right) = F(0)\frac{\Lambda_1^2}{Q^2-\Lambda_1^2}\frac{\Lambda_2^2}{Q^2-\Lambda_2^2}.
\end{equation}
By fitting Eq.~\eqref{eq:FQlambda} with Eq.~\eqref{eq:FQzeta}, we can obtain the values of $\Lambda_1$ and $\Lambda_2$ for each form factor, which are listed in Table~\ref{Tab:L1L2}.

\begin{figure}[t]
	\centering	\includegraphics[width=1.0\linewidth]{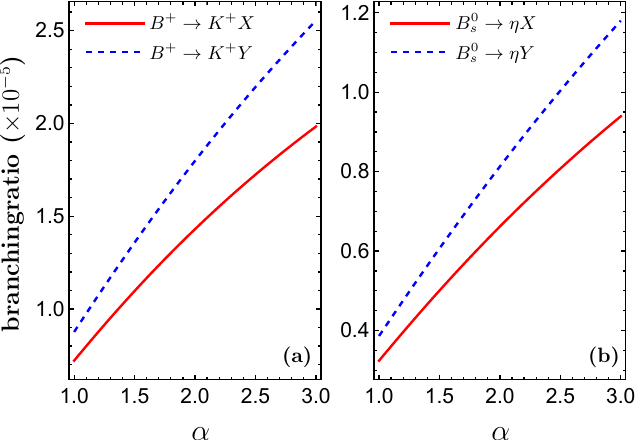}
	\caption{Branching fractions of $B^+ \rightarrow K^+X/Y$(diagram (a)) and $B_s^0 \rightarrow \eta X/Y$ (diagram(b)) depending on the parameter $\alpha$.}
	\label{Fig:BR}
\end{figure}

\subsection{Branching Fractions}

Now all the relevant coupling constants have been determined except model parameter $\alpha$. As indicated in Ref.~\cite{Cheng:2004ru}, the model parameter $\alpha$ should be of order unity. In the present work, we vary $\alpha$ from 1 to 3 to check the parameter dependences of the estimated branching fractions. In Fig.~\ref{Fig:BR}, we present our estimations of the branching fractions of $B^+ \to K^+ X(4630)/Y(4626)$ (diagram (a)) and $B_s^0 \to \eta X(4630)/Y(4626)$ (diagram (b)) depending on the model parameter $\alpha$. In the figure we can find that the branching fractions of $B^+ \to K^+ X(4630)$ is estimated to be of order $10^{-5}$, and increase with the increasing of parameter $\alpha$. In particular, in the considered parameter range, the branching ratio is estimated to be $\left(1.43^{+0.55}_{-0.71}\right) \times 10^{-5}$, where the central value is estimated with $\alpha=2$, while the uncertainties are caused by the variation of $\alpha$ from $1$ to $3$. As indicated in Refs.~\cite{LHCb:2021uow,ParticleDataGroup:2020ssz}, the branching fraction of $B^+ \to K^+ X(4630) \to K^+ J/\psi \phi$ is estimated to be $\left(1.3^{+1.5}_{-0.8}\right) \times 10^{-6}$. Then we can conclude that if one intend to understand the measured branching fraction of $B^+ \to K^+ X(4630) \to K^+ J/\psi \phi$ in the $D_s^{\ast +} {D}_{s1}(2536)^-$ molecular frame, the branching fraction of $X(4630)\to J/\psi \phi$ should be $\left(9.09^{+10.4}_{-7.19}\right)\%$. In general, the branching fractions of the hidden charm decay processes for the tetraquark or pentaquark molecular states is of several percent, which is consistent with expectation of the present estimations. 

Besides, the process $B^+ \to K^+ X(4630)$, the branching fraction of $B^+ \to K^+ Y(4260)$ is also estimated as shown in Fig.~\ref{Fig:BR}-(a). In the considered parameter range, the branching fraction of $B^+ \to K^+ Y(4260)$ is evaluated as $\left(1.80^{+0.75}_{-0.92}\right) \times 10^{-5} $. Moreover, we also investigate the productions of $X(4630)$ and $Y(4626)$ in the $B_s^0$ decays along with a $\eta$ meson. The $\alpha$ dependence of the branching fractions are presented in Fig.~\ref{Fig:BR}-(b). From the figure, one can find that the $\alpha$ dependence of the branching fractions of $B_s^0 \to \eta X(4630)/Y(4626)$ are very similar to that of $B^+ \to K^+ X(4630)/Y(4626)$. In particular, the branching fractions are estimated to be $\left(0.66^{+0.28}_{-0.34}\right) \times 10^{-5} $ and $\left(0.81^{+0.37}_{-0.42}\right) \times 10^{-5}$ for $B_s^0 \to \eta X(4630)$ and $B_s^0 \to \eta Y(4626)$, respectively.

\begin{figure}[t]
	\centering
	\includegraphics[width=8.5cm]{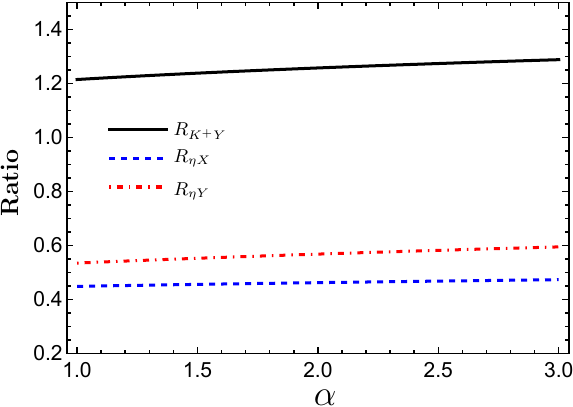}
	\caption{The ratios of the branch fractions of different channels to the branch fraction of $B^+ \to K^+X$ depending on the parameter $\alpha$.}
	\label{Fig:Ratio}
\end{figure}

From Fig.~\ref{Fig:BR}, we find the $\alpha$ dependences of the relevant branching fractions are similar, thus we further defined the ratios of the branching fractions as, 
\begin{eqnarray}
	R_{K^+Y}=\frac{B^+ \to K^+ Y(4626)}{B^+ \to K^+ X(4630)}, \nonumber\\
	R_{\eta X}=\frac{B_s^0 \to \eta X(4630)}{B^+ \to K^+ X(4630)}, \nonumber\\
    R_{\eta Y}=\frac{B_s^0 \to \eta Y(4626)}{B^+ \to K^+ X(4630)},
\end{eqnarray}
which is expected to weakly dependent on the model parameter $\alpha$. The $\alpha$ dependence of the ratios are presented in Fig.~\ref{Fig:Ratio}. Our estimations indicate that $R_{K^+Y}=1.22\sim 1.29$, $R_{\eta X}=0.45\sim 0.47$, $R_{\eta Y}=0.54\sim 0.59$. 
Combining the $R_{\eta X}$ estimated in the present work and the measured branching ratio of $B^+\to K^+ X(4630)\to K^+J/\psi \phi$ by the LHCb Collaboration, one can conclude,
\begin{eqnarray}
\mathcal{B}[B_s^0 \to \eta X(4630) \to \eta J/\psi \phi] =\left(6.0^{6.9}_{-3.7}\right) \times 10^{-7}.
\end{eqnarray}
The above predictions of the ratios and the branching ratio of $B_s^0 \to \eta X(4630) \to \eta J/\psi \phi$ are almost independent on the model parameter, indicating that they could serve as a crucial test for the molecular interpretations of $X(4630)$ and $Y(4626)$.

\begin{figure}[t]
	\centering
	\includegraphics[width=8.5cm]{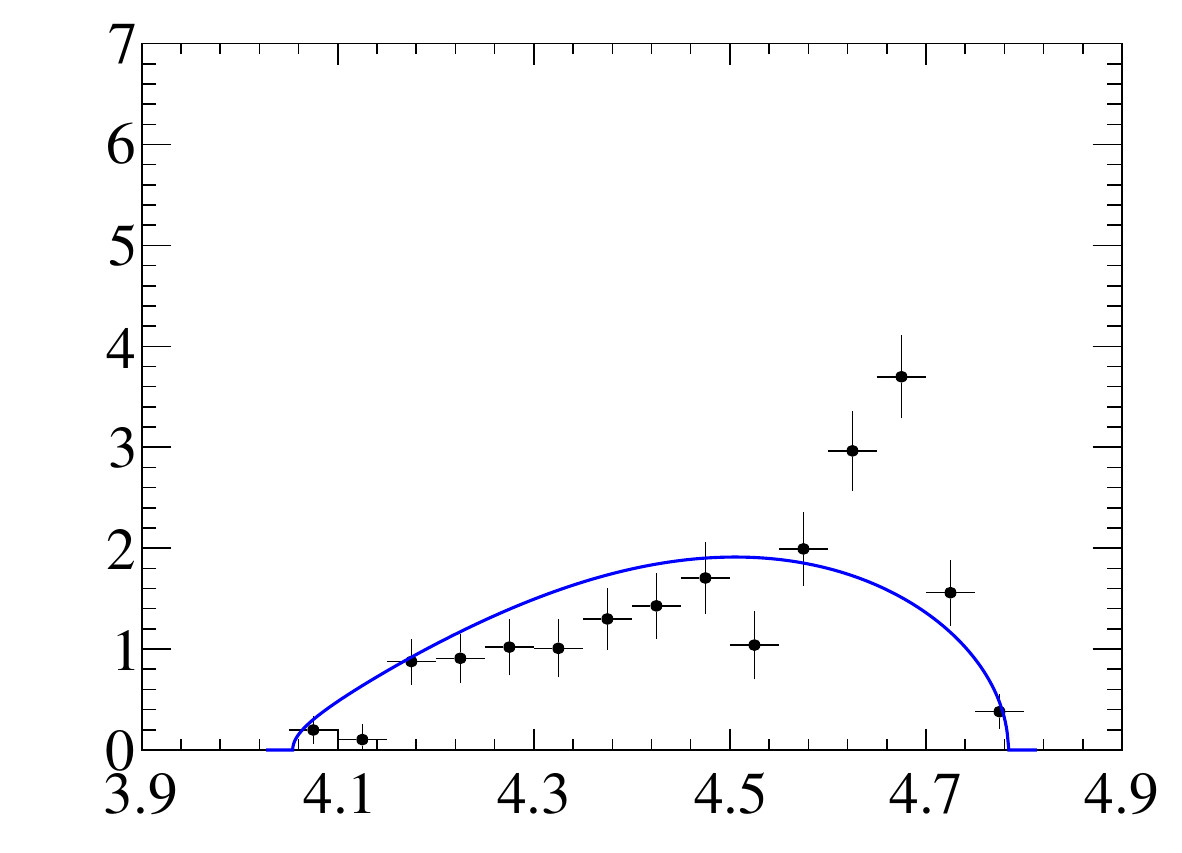}
	\put(-130,-5){\large$m_{J/\psi \eta^\prime}$}
	\put(-245,50){\rotatebox{90}{\Large $\frac{1}{N}\frac{dN}{dm_{J/\psi \eta^\prime}}$[$1\over\mathrm{GeV}^{-2}$]}}
	\caption{The $J/\psi \eta^\prime$ invariant mass distribution of $B^+ \to K^+ J/\psi \eta^\prime$ reported by LHCb Collaboration~\cite{LHCb:2023qca}. The blue solid curve is the corresponding phase space.}
	\label{Fig:LHCb}
\end{figure}

Before the end of this work, it is worth mentioning that in Ref.~\cite{LHCb:2023qca}, the LHCb Collaboration reported the analysis of the $B^+ \to J/\psi \eta^\prime K^+ $ by using the data corresponding to a total integrated luminosity of $9\ \mathrm{fb}^{-1}$. The $J/\psi \eta^\prime $ invariant mass distributions were measured. For the convenience of readers, we present the $J/\psi \eta^\prime $ invariant mass distributions in Fig.~\ref{Fig:LHCb}, where one can find that a significant structure is observed in the high mass region, which may include the contributions from $Y(4626)$. Further full amplitude analysis with larger data sample and low background levels may provide an opportunity of searching $Y(4626)$ in $B^+\to K^+ J/\psi \eta^\prime$. Moreover, the branching fraction of $B^+ \to K^+ J/\psi \eta$ is $(1.24\pm 0.14)\times 10^{-4}$, which is about four times of the one for $B^+ \to K^+ J/\psi \eta^\prime$. Thus, searching $Y(4626)$ in the process $B^+ \to K^+ J/\psi \eta$ may be also accessible.

\section{SUMMARY}
In the year 2021, the LHCb Collaboration reported a full amplitudes analysis of the $B^+ \to K^+ J/\psi \phi$, seven charmonium-like states were observed in the $J/\psi \phi $ invariant mass distributions, among which $X(4630)$ with $J^{PC}=1^{-+}$ is particularly fascinating, since it could not be a conventional meson states. It is interesting to note that the exotic state $X(4630)$ has a 'pigeon pair' partner with almost the same mass but different $C$-parity, named $Y(4626)$, which was observed in the cross sections for $e^+e^- \to D_s^{\ast +} D_{s1}^-+c.c.$ in 2019. Both states are very close to the threshold of $D_{s}^{\ast +} D_{s1}(2536)^-$, which indicates that $x(4630)$ and $Y(4626)$ could be good molecular candidates composed of $D_{s}^{\ast +} D_{s1}(2536)^-$.

Considering the observed process of $X(4630)$, we investigate the production processes $B^+\to K^+ X(4630)/Y(4626)$ in the $D_s^{\ast +} D_{s1}(2536)^-$ molecular frame in the present work, our estimations indicate that the branching fraction of $B^+ \to K^+ X(4630)$ is estimated to be of order $10^{-5}$, which is consistent with the experimental expectations. In addition, the branching fraction of $B^+ \to K^+ Y(4626)$ is about $1.2\sim 1.3$ times that of $B^+\to K^+ X(4630)$, indicating that the process $B^+ \to K^+ Y(4626)$ should be potentially observed experimentally, and we propose to search $Y(4626)$ in the process $B^+ \to K^+ J/\psi \eta^{(\prime)}$. Moreover, we also investigate the production of $X(4630)$ and $Y(4626)$ in the $B_s^0$ decay processes, and the branching fractions are estimated to be about half of the one of $B^+ \to K^+X(4630)$.

\section*{ACKNOWLEDGMENTS}
The author ZY would like to thank Pei-Dong Yan for his useful help. This work is supported by the National Natural Science Foundation of China under the Grant Nos. 12175037 and 12335001.
	
\bibliographystyle{unsrt}
\bibliography{4630.bib}
\nocite{*}
\end{document}